\documentclass[onecolumn]{aastex}

\usepackage{amsmath}
\usepackage[usenames,dvipsnames]{color}
\usepackage{amssymb}

\newcommand{\omegabold}{\mbox{\boldmath$\omega$}}

\slugcomment{Submitted to the Astrophysical Journal}

\shorttitle{}
\shortauthors{Mart\' inez Gonz\'alez \& Asensio Ramos}

\begin{document}

\title{Analytical calculation of Stokes profiles of rotating stellar magnetic dipole}
\author{M. J. Mart\'{\i}nez Gonz\'alez \& A. Asensio Ramos}
\affil{Instituto de Astrof\' isica de Canarias, V\' ia L\'actea s/n, 38200, La
Laguna, Tenerife, Spain}
\affil{Departamento de Astrof\'{\i}sica, Universidad de La Laguna, E-38205 La Laguna, Tenerife, Spain}

\begin{abstract}
The observation of the polarization emerging from a rotating star at different phases opens up
the possibility to map the magnetic field in the stellar surface thanks to the well-known Zeeman
Doppler Imaging. When the magnetic field is sufficiently weak, the circular and linear polarization
profiles locally in each point of the star are proportional to the first and second derivatives of 
the unperturbed intensity profile, respectively. We show that the weak-field approximation (for weak lines in
the case of linear polarization) can
be generalized to the case of a rotating star including the Doppler effect and taking into account 
the integration on the stellar surface. The Stokes profiles are written as a linear
combination of wavelength-dependent terms expressed as series expansions in terms of Hermite
polynomials. These terms contain the surface integrated magnetic field
and velocity components. The direct numerical evaluation of these quantities is limited to rotation velocities
not larger than 8 times the Doppler width of the local absorption profiles.
Additionally, we demonstrate that, in a rotating star, the circular polarization flux depends 
on the derivative of the intensity flux with respect to the wavelength and also on the profile itself. 
Likewise, the linear polarization depends on the profile and on its first and second derivative with 
respect to the wavelength. We particularize the general expressions to a rotating dipole.
\end{abstract}


\section{Introduction}

The present knowledge of solar and stellar magnetism derives essentially from
the interpretation of polarization in spectral lines due to 
the Zeeman effect\footnote{There are some notable exceptions as, for example, the 
study of the quiet Sun magnetism by means of the interpretation of scattering polarization 
and its modification by the Hanle effect \citep[e.g.,][for a recent review]{trujillo_spw6_11}}. 
In the presence of a weak magnetic field, the 
radiative transfer equation for polarized light can be solved analytically
and it is known as the ``weak field approximation''. Although very simple, it relies
on a set of strong assumptions that we discuss later. However, the weak field approximation is applicable in 
many scenarios and gives very good results for the inference of magnetic fields
as compared with more elaborate methods. In solar physics, it is at the heart of
the success of a large number of synoptic magnetographs, like those of Big Bear \citep{varsik95,spirock01}. It
has been also used recently to produce vector magnetograms with the 
Imaging Magnetograph Experiment (IMaX) instrument \citep{imax11}
onboard the Sunrise balloon \citep{sunrise10}. In stellar spectropolarimetry, the weak field
approximation is at the base of the least-squares deconvolution \citep[LSD;][]{donati97}, 
the most successful technique used to detect and measure magnetic fields in solar-type 
stars, or in some recent works on central stars of planetary nebulae \citep{jordan05,leone11}, 
white dwarfs \citep{aznar_cuadrado04}, pulsating stars \citep{silvester09}, hot subdwarfs \citep{otoole05} and
Ap and Bp stars \citep{wade00,bagnulo02}.

Most of the scientific cases commented so far deal with unresolved structures
which means that the weak field approximation has to be 
understood in terms of fluxes of the Stokes parameters instead of specific
intensities. Recently, \cite{marian_dipolo12} derived the weak field expressions in terms of
fluxes, particularizing to the case of the stellar dipole. They neglected rotation to
simplify the equations and to derive analytical expressions for the 
inference of the magnetic field. When rotation is taken into account (in other words,
there is a correlation between the magnetic field and the line-of-sight velocity), 
the problem becomes much more challenging and it is usually solved numerically
\citep[e.g.,][for a recent effort]{petit_wade12}. However, we show in this paper that
it is possible to obtain analytical expressions for the Stokes vector fluxes in the 
weak field approximation. There are two interests on this effort. First,
having an analytical expression for the Stokes flux might be of help to investigate
the interplay between rotation and the magnetic field. Second, it can be used to speed up inversion codes
based on the weak-field approximation because no numerical integration is needed. This will facilitate the application of
Bayesian inference codes, like the one developed by \cite{petit_wade12}, that are based
on Markov Chain Monte Carlo methods that need to carry out the synthesis thousands of times
to sample the posterior distribution.

This is the first paper of a series dealing with analytical expressions for the fluxes 
of the Stokes parameters in the weak field limit. This paper deals with
the expressions for a rotating star with a magnetic field on its surface. The
results depend on the correlation 
between the magnetic field and the velocity and are general for any rotation profile
or any magnetic field configuration. We particularize the equations to the case of a 
rotating magnetic stellar dipole.

\section{The weak field approximation for a star}
\begin{figure}[!t]
\centering
\includegraphics[width=0.5\textwidth]{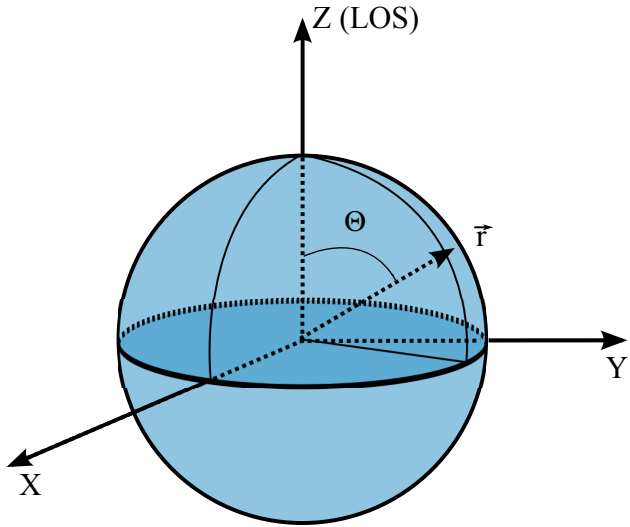}
\caption{This figure shows the reference systems used to define the Stokes parameters in this
work. We consider that the observer is along the $Z$ axis. The direction $X$ defines the reference 
for $Q>0$. The vector $\mathbf{r}$ defines any position on the stellar surface.}
\label{fig:fig_geom}
\end{figure}

Let us consider a star with arbitrary velocity and magnetic fields on its surface.
We define both fields for an arbitrary point $\mathbf{r}$ in the surface using the 
reference frame represented in Fig. \ref{fig:fig_geom}. The line-of-sight (LOS) is 
along the $Z$ axis, while we choose the axis $X$ as the one defining the reference direction for positive Stokes $Q$.

Our aim is to compute the local Stokes profiles emerging from every point
of the visible surface and compute its integral. As explained in the introduction,
a simple (although subject to approximations) way to relate the emergent Stokes profiles with the magnetic field
vector is the weak-field approximation. This approximation (combined with the weak-line approximation for
linear polarization)
leads to an analytical solution to the radiative transfer
equation. It is based on the following three fundamental assumptions: i) the magnetic field vector 
is constant along the line of sight on the region of the stellar atmosphere where the spectral line 
is formed, ii) the magnetic field intensity is sufficiently weak so that the
Zeeman splitting is smaller than any other line broadening mechanism and iii) 
the LOS velocity and any broadening mechanism are
constant with height in the line formation region. 
Under these simplifying conditions, the magnetic field can be considered as a perturbation to the
zero-field case. This approximation is not strictly valid in general, but it is a
good approximation whenever the magnetic field strength is weak in the line 
formation region in the absence of strong gradients along the LOS.
Quantitatively, the approximation holds whenever 
$g \Delta \lambda_B\ll\Delta \lambda_D$, where $\Delta \lambda_B$ \citep[see Eq. (3.14) in][]{landi_landolfi04}
is the Zeeman splitting that is proportional to the field strength, $\Delta \lambda_D$ is the dominant broadening mechanism
and $g$ represents the effective Land\'e factor of the line, which
quantifies the magnetic sensitivity of the line and that only depends on the quantum numbers of 
the transition \citep[see, e.g.,][]{landi_landolfi04}. 
From its definition, it is clear that the weak field regime occurs at different
field strengths for different spectral lines, for different 
stellar objects and for different spectral resolutions. For instance, for a magnetically 
sensitive iron spectral line with $g=3$ at 5000 \AA\ with a standard thermal broadening at $T=6500$ K, the 
weak-field regime is applicable for fields below $\sim$600 G if the spectral resolution of the
spectrograph is smaller than the line broadening. 

Under the previous simplifying assumptions, and to first order in $\Delta \lambda_B$, the intensity 
profile ${I}(\lambda)$ of a spectral line
formed in a weak magnetic field is fully insensitive to the magnetic field. 
In other words, it fulfills the standard radiative transfer equation in the absence of a magnetic
field \citep[e.g.,][]{mihalas78}. Likewise, at first order in $\Delta \lambda_B$, the circular polarization profile, i.e., 
the Stokes ${V}(\lambda)$ profile for a given spectral line, has the following
expression:
\begin{equation}
{V}(\lambda)=-{C}\lambda_0^2 g B_z \frac{\partial {I}(\lambda)}{\partial \lambda},
\label{eq:v}
\end{equation}
where the symbol $B_{z}$ stands for the longitudinal component of the magnetic
field (the projection along the LOS in the reference system shown in Fig. \ref{fig:fig_geom}). 
The symbol $\lambda$ stands for the wavelength while $\lambda_0$ is the central
wavelength of the transition. The constant $C=4.67\times 10^{-13} \mathrm{G}^{-1}
\mathrm{\AA}^{-1}$.

At first order in $\Delta \lambda_B$, linear polarization is zero. In order
to obtain an expression for the Stokes profiles characterizing 
linear polarization we have to expand the radiative transfer equation to second
order in $\Delta \lambda_B$ and assume that the spectral line is not 
saturated and weak. Under these assumptions, a good approximation to the intensity profile emerging from
the stellar atmosphere is to consider it, in general, proportional to a Voigt function \citep[see Eqs. (9.84) and (9.86) in][]{landi_landolfi04}. 
For the sake of simplicity, we consider the case of a Gaussian line profile. All our results can be
derived for a Voigt profile but the final expressions are more complicated without any fundamental
difference.
This assumption somehow reduces the scope of applicability of our results for linear polarization because the line
shape self-similarity is only valid for lines whose line depth is smaller than $\sim$40\% according
to the results of \cite{kochukhov10}.

With this assumption, the weak-field approximation
for the Stokes ${Q}(\lambda)$ and ${U}(\lambda)$, defined in the $XYZ$
reference system shown in Fig. \ref{fig:fig_geom}, are \citep[see][]{landi_landolfi04,marian_dipolo12}:
\begin{eqnarray}
{Q}(\lambda)&=&-\frac{{C}^2}{4}\lambda_0^4 G (B_{x}^2 - B_{y}^2)
\frac{\partial^2 {I}(\lambda)}{\partial \lambda^2} \nonumber \\
{U}(\lambda)&=&-\frac{{C}^2}{2}\lambda_0^4 G B_{x}B_{y} \frac{\partial^2
{I}(\lambda)}{\partial \lambda^2},
\label{eq:q_u}
\end{eqnarray}
where the symbol $G$ plays the role of the effective Land\'e factor for linear
polarization and quantifies the sensitivity of 
linear polarization to the magnetic field. Again, it is only a function of the
quantum numbers of the transition \citep[][]{landi_landolfi04}.
The symbols $B_{x}$ and $B_{y}$ stand for the components of the magnetic field vector along the
$X$ and $Y$ axes, respectively. The Stokes profiles in any other reference system that is defined as a rotation
around the $Z$ axis by an angle $\gamma$ are given by the usual rotation \citep{landi_landolfi04}:
\begin{eqnarray}
{Q}_\mathrm{new}(\lambda) &=& \cos (2\gamma) {Q}(\lambda) + \sin (2\gamma) {U}(\lambda) \nonumber \\
{U}_\mathrm{new}(\lambda) &=& -\sin (2\gamma) {Q}(\lambda) + \cos (2\gamma) {U}(\lambda),
\label{eq:rotation_stokes}
\end{eqnarray}
where the quantities with the subindex ``new'' refer in this case to the new reference frame.

When observing unresolved objects, the detected polarized radiation is obtained
as an integration in the plane of the sky of the 
individual Stokes parameters at each point of the stellar surface ($\Sigma$), all of them
referred to a common reference system. We describe
the stellar disk using the polar coordinates $\rho$ and $\theta$, where $\rho$ is
normalized to the stellar radius, so that its range is from 0 to 1 \citep[see][]{marian_dipolo12}. Therefore,
the average of any arbitrary function $h(\rho,\theta)$ on the surface $\Sigma$ will be given by:
\begin{equation}
\overline{h}=\int \int d\Sigma h(\rho,\theta)= \int_0^1 \rho d\rho \int_0^{2\pi} d\theta
h(\rho,\theta).
\label{eq:integration_sky}
\end{equation}
As a consequence, it is obvious that the observed values of the Stokes flux vector
depends on the surface distribution of the magnetic field, on the centre-to-limb variation (CLV) of the
radiation, and on the Doppler effect due to the rotation of the star. 

Following standard approximations and assuming that the star has azimuthal symmetry, \cite{marian_dipolo12} 
parameterized the intensity at each point of the star as a 
product of two functions: $I_0(\lambda)$, that only depends on the wavelength (the intensity at the
disc centre since they neglect rotation), and $f(\mu)$, the CLV that depends on the position on the
disk and is wavelength independent \citep[e.g.,][]{claret00,cox00}:
\begin{equation}
f(\mu)=(1-a-b+a\mu+b\mu^2).
\label{eq:clv}
\end{equation}
The CLV is given in terms of $\mu=\cos \Theta=\sqrt{1-\rho^2}$, where $\Theta$
is the astrocentric angle between the normal to a point in the stellar surface
and the line of sight (see Fig. \ref{fig:fig_geom}). 
The parameters $a$ and $b$ have values between 0 and 1 and are supposed to
be constant along the spectral line (although they can vary from line to
line). With this assumption, we also force the line depth
to be $\mu$-independent. Note that the values of $a$ and $b$ have to fulfill the condition 
that the intensity has to be positive. This simple parameterization allowed them
to express the weak field 
equations in terms of the observed fluxes, thus mimicking the expressions for the
specific Stokes vector. This was possible because the intensity flux is assumed to be
the product of the wavelength-dependent intensity at disk center and the wavelength
independent CLV. Using this relationship, the
weak-field and weak-line approximation for the integrated polarized flux remains formally the
same than for the specific Stokes vector but the 
components of the magnetic field appear weighted by the CLV law. 

When the velocity field is included, the situation turns out to be much more complicated because the
intensity itself depends on the local projection of the velocity along the
$Z$ axis ($v_z$), which subsequently depends on the position along the stellar disk in a potentially
complicated way. In this case, the fluxes of the Stokes vector are expressed as:
\begin{eqnarray}
F_{I}(\lambda)&=&\int_0^1 \rho d\rho \int_0^{2\pi} d\theta~{I}(\lambda, v_z(\rho,\theta),\mu) \nonumber \\
F_{Q}(\lambda)&=&-\frac{{C}^2}{4}\lambda_0^4G \int_0^1 \rho d\rho \int_0^{2\pi} d\theta~ 
\left[B_x^2(\rho,\theta)-B_y^2(\rho,\theta) \right]\frac{\partial^2 {I}(\lambda, v_z(\rho,\theta),\mu)}{\partial
\lambda^2} \nonumber \\
F_{U}(\lambda)&=&-\frac{{C}^2}{2}\lambda_0^4G \int_0^1 \rho d\rho \int_0^{2\pi} d\theta~B_x(\rho,\theta)B_y(\rho,\theta)
\frac{\partial^2 {I}(\lambda, v_z(\rho,\theta),\mu)}{\partial \lambda^2} \nonumber \\
F_{V}(\lambda)&=&-{C}\lambda_0^2g \int_0^1 \rho d\rho \int_0^{2\pi} d\theta~B_z(\rho,\theta) \frac{\partial
{I}(\lambda,v_z(\rho,\theta),\mu)}{\partial \lambda},
\end{eqnarray}

The immediate consequence of the dependence on $v_z(\rho,\theta)$ is that the integrals cannot
be separated. To proceed, it is crucial to decouple the rotation velocity and the wavelength to end up
with expressions for the weak field approximation in terms of the observed fluxes. In order to do
that, we assume that the shape of the local absorption profile is independent of the position 
on the stellar surface, apart from the Doppler shift introduced by $v_z$, and that the CLV is
wavelength independent. Furthermore, if we assume that the absorption profile is
Gaussian, we end up with:
\begin{equation}
{I}(\lambda, v_z(\rho,\theta), \mu)= f(\mu) I(\lambda,v_z(\rho,\theta),\mu=1) = f(\mu)\left[I_c-\eta
\exp\left[-\frac{1}{2}\left(\frac{\lambda-\lambda_0(1+v_z(\rho,\theta)/c)}{\sigma}
\right)^2\right]\right],
\label{eq:intensity_gaussian}
\end{equation}
where $I_c$ stands for the continuum intensity, $\eta$ for the line depth, and
$\sigma$ for the line width. This approximation is not appropriate for stars with
large radial temperature and density gradients because the lines close
to the limb will differ strongly from the lines at the disk center.
Apart from the Gaussian shape, it is possible to work with Voigt functions but the 
subsequent expressions turn out to be more complicated and no new physics is gained,
so we prefer to stick with the Gaussian shape. The
Gaussian shape is also consistent with the assumption of an unsaturated and weak spectral
line that is necessary for Eqs. (\ref{eq:q_u}) to hold.

We can carry out the Taylor
expansion of $I(\lambda,v_z,\mu=1)$ around $v_z=0$ (we drop now the $\mu$-dependence) to obtain the following
expression for the intensity at each point of the stellar surface:
\begin{equation}
I(\lambda, v_z)=I(\lambda,0)+\sum_{n=1}^\infty \frac{1}{n!} \frac{\partial^n
I(\lambda, v_z)}{\partial v_z}\bigg|_{v_z=0} v_z^n.
\end{equation}
Defining the emission line profile at zero velocity as
\begin{equation}
\psi(\lambda)=\eta
\exp\left[-\frac{1}{2}\left(\frac{\lambda-\lambda_0}{\sigma}\right)^2\right],
\end{equation}
the absorption profile at an arbitrary velocity shift can be written as:
\begin{equation}
I(\lambda, v_z)=I_c - \psi(\lambda) +\sum_{n=1}^\infty \frac{1}{n!} \frac{\partial^n
I(\lambda, v_z)}{\partial v_z}\bigg|_{v_z=0} v_z^n.
\end{equation}

Taking into account that the Gaussian function is the generating
function of the Hermite polynomials:
\begin{equation}
\frac{d^n}{dx^n}\exp\left[-\frac{x^2}{2}\right]=(-1)^n\exp\left[-\frac{x^2}{2}
\right] \textit{He}_n(x),
\end{equation}
where $\textit{He}_n(x)$ is the Hermite polynomial of order $n$ with weight function $w(x)=\exp(-x^2/2)$,
also known as Chebyshev-Hermite polynomials \citep[see][]{abramowitz72}. If a Voigt function is used
instead for the local profile, the derivatives are more complex and the recurrence relations
of \cite{heinzel78} can be used. Taking the previous definition into account, 
the derivatives ($n>0$) of the intensity profile can be expressed as:
\begin{equation}
\frac{\partial^n I(\lambda, v_z)}{\partial
v_z^n}\bigg|_{v_z=0}=-\left(\frac{\lambda_0}{\sigma c}\right)^n \psi(\lambda)
\textit{He}_n \left( \frac{\lambda-\lambda_0}{\sigma} \right),
\end{equation}
so that the intensity profile of Eq. (\ref{eq:intensity_gaussian}) is given by the
following Gram-Charlier series:
\begin{equation}
{I}(\lambda, v_z, \mu)=f(\mu)\left[ I_c - \psi(\lambda)\sum_{n=0}^\infty
\frac{1}{n!}\left(\frac{\lambda_0}{\sigma c}\right)^n
\textit{He}_n \left( \frac{\lambda-\lambda_0}{\sigma} \right) v_z^n  \right],
\end{equation}
where we have extended the summation to $n=0$ to include the profile at zero velocity, since $\textit{He}_0(x)=1$.
The previous formula shows that the intensity is given as the addition
of infinite terms (potentially a large number of them for obtaining convergence) given by the product of a function that
depends on the wavelength and another one that depends on the velocity along the
line of sight. The latter is the only one that changes on the stellar surface, so
that this variable separation allows us to carry out the surface integration. In order to simplify the following 
computations, it is interesting to define the function 
\begin{equation}
\mathcal{H}(\lambda, v_z)=\sum_{n=0}^\infty
\frac{1}{n!}\left(\frac{\lambda_0}{\sigma c}\right)^n
\textit{He}_n \left( \frac{\lambda-\lambda_0}{\sigma} \right) v_z^n,
\end{equation}
so that the intensity profile and its first and second derivatives with respect to the
wavelength are given by:
\begin{eqnarray}
{I}(\lambda, v_z, \mu) &=&f(\mu) \left[ I_c-\psi(\lambda)\mathcal{H}(\lambda,v_z) \right] \nonumber \\
{I}'(\lambda, v_z, \mu)&=& -\psi'(\lambda) f(\mu) \mathcal{H}(\lambda,v_z) - \psi(\lambda) f(\mu) \mathcal{H}'(\lambda,v_z) \nonumber \\
{I}''(\lambda, v_z, \mu)&=& -\psi''(\lambda) f(\mu) \mathcal{H}(\lambda,v_z) - 2\psi'(\lambda) f(\mu) \mathcal{H}'(\lambda,v_z)
-\psi(\lambda) f(\mu) \mathcal{H}''(\lambda,v_z),
\end{eqnarray}
where the $'$ and $''$ symbols represent the first and second
derivatives with respect to the wavelength.

Applying the previous developments and dropping the dependences of the
variables to avoid crowding, the polarized flux in the weak-field weak-line approximation can be
written as:
\begin{eqnarray}
F_{I}(\lambda) &=&
F_c-\psi\overline{f\mathcal{H}} \nonumber \\
F_{Q}(\lambda) &=& \frac{{C}^2}{4}\lambda_0^4G 
\left[ \psi''(\lambda) \overline{(B_x^2-B_y^2)f\mathcal{H}} + 2\psi'(\lambda) \overline{(B_x^2-B_y^2)f\mathcal{H}'} +
\psi(\lambda) \overline{(B_x^2-B_y^2)f\mathcal{H}''} \right] \nonumber \\
F_{U}(\lambda) &=& \frac{{C}^2}{2}\lambda_0^4G 
\left[ \psi''(\lambda) \overline{B_xB_yf\mathcal{H}} + 2\psi'(\lambda) \overline{B_xB_yf\mathcal{H}'} +\psi(\lambda) \overline{B_xB_yf\mathcal{H}''} \right] \nonumber \\
F_{V}(\lambda)&=&{C}\lambda_0^2g \left[ \psi'(\lambda) \overline{B_zf\mathcal{H}} +\psi(\lambda) \overline{B_zf\mathcal{H}'} \right],
\label{eq:fI_star}
\end{eqnarray}
where we remind that the overlines represent integration in the plane of the sky,
as indicated in Eq. (\ref{eq:integration_sky}). It is important to note that the
area asymmetry of the polarized flux is zero as can be verified by computing the
wavelength integral of the previous equations (it is easy to verify that the integral
is zero by using the properties of the Hermite polynomials).

The symbol $F_c$ stands for the continuum intensity flux and it 
is trivially given by:
\begin{equation}
F_c = \pi I_c \left(1-\frac{a}{3}-\frac{b}{3} \right).
\end{equation}
As expected, the Stokes parameters depend on
quantities averaged over the stellar surface. Using the expansion on Hermite
polynomials of the absorption spectral line, the terms appearing in the previous equations 
can be written as the following summations:
\begin{eqnarray}
\label{eq:averages_surface}
\overline{B_z f \mathcal{H}} &=& -\frac{1}{2} \sum_{n=0}^\infty
\frac{1}{n!} \left( \frac{\lambda_0}{\sigma c} \right)^n 
\textit{He}_n\left(\frac{\lambda-\lambda_0}{\sigma}\right) X_1(n) \nonumber \\
\overline{(B_x^2-B_y^2) f \mathcal{H}} &=& \frac{1}{4}
\sum_{n=0}^\infty \frac{1}{n!} \left( \frac{\lambda_0}{\sigma c}
\right)^n \textit{He}_n\left(\frac{\lambda-\lambda_0}{\sigma}\right) X_2(n) \nonumber \\
\overline{B_x B_y f \mathcal{H}} &=& \frac{1}{4} \sum_{n=0}^\infty
\frac{1}{n!} \left( \frac{\lambda_0}{\sigma c} \right)^n 
\textit{He}_n\left(\frac{\lambda-\lambda_0}{\sigma}\right) X_3(n) \nonumber \\
\overline{f \mathcal{H}} &=& \sum_{n=0}^\infty \frac{1}{n!} \left(
\frac{\lambda_0}{\sigma c} \right)^n 
\textit{He}_n\left(\frac{\lambda-\lambda_0}{\sigma}\right) X_4(n).
\label{eq:averages}
\end{eqnarray}
The functions $X_i(n)$ are given by:
\begin{eqnarray}
X_1(n) &=& \int_0^1 \rho d\rho \int_0^{2\pi} d\theta f(\mu) B_z(\rho,\theta) v_z^n(\rho,\theta) \nonumber \\
X_2(n) &=& \int_0^1 \rho d\rho \int_0^{2\pi} d\theta f(\mu) (B_x^2(\rho,\theta)-B_y^2(\rho,\theta)) v_z^n(\rho,\theta) \nonumber \\
X_3(n) &=& \int_0^1 \rho d\rho \int_0^{2\pi} d\theta f(\mu) B_x(\rho,\theta) B_y(\rho,\theta) v_z^n(\rho,\theta) \nonumber \\
X_4(n) &=& \int_0^1 \rho d\rho \int_0^{2\pi} d\theta f(\mu) v_z^n(\rho,\theta).
\end{eqnarray}

It is important to note that Eqs. (\ref{eq:fI_star}) and (\ref{eq:averages_surface}) are formally invariant for any magnetic field
configuration, rotation profile and center-to-limb variation once the assumptions stated before (sufficiently weak 
magnetic field, wavelength independent CLV, unsaturated and weak Gaussian line) are fulfilled. The information
about the magnetic configuration, rotation and the CLV is encoded in the $X_i(n)$ functions, which may depend on 
additional external parameters. This is the case of the dipolar field with solid body rotation (see Sec. \ref{sec:dipole}), 
in which these functions depend on the orientation of the dipole and the angular velocity vector.

The integrated Stokes parameters also depend on terms that include the
wavelength derivative of the
$\mathcal{H}$ function. Thanks to the effective separation between
wavelength and velocity, these
are easily obtained from the previous ones by using known properties
of the Hermite polynomials. For instance:
\begin{eqnarray}
\label{eq:averages_surface_d}
\overline{B_z f \mathcal{H}'} = -\frac{1}{2} \sum_{n=1}^\infty
\frac{1}{(n-1)!} \frac{1}{\sigma} \left( \frac{\lambda_0}{\sigma c}
\right)^n 
\textit{He}_{n-1}\left(\frac{\lambda-\lambda_0}{\sigma}\right) X_1(n) \nonumber \\
\overline{B_z f \mathcal{H}''} = -\frac{1}{2} \sum_{n=2}^\infty
\frac{1}{(n-2)!} \frac{1}{\sigma^2} \left( \frac{\lambda_0}{\sigma c}
\right)^n 
\textit{He}_{n-2}\left(\frac{\lambda-\lambda_0}{\sigma}\right) X_1(n),
\end{eqnarray}
and the rest of terms are computed following the same strategy.

When the star is not rotating, the only terms contributing to the summations in Eq. (\ref{eq:averages_surface})
are those with $n=0$. Likewise, the terms with derivatives of Eqs. (\ref{eq:averages_surface_d}) 
are always zero.

\section{The weak-field approximation for observers}
The polarized fluxes of Eqs. (\ref{eq:fI_star}) can be used to explain stellar observations using
a standard optimization method to fit the functional forms to the observations. 
However, we note that the expressions for the polarized fluxes depend on $\eta$ and $\sigma$, 
properties of the local line profiles to which we do not have direct access. It is possible, though, to 
drop the dependence on $\eta$ from the equations if we use the fact that $\psi(\lambda)$, $\psi'(\lambda)$ and
$\psi''(\lambda)$ can be written in terms of $F_{I}(\lambda)$, $F_{I}'(\lambda)$
and $F_{I}''(\lambda)$, which are, otherwise, observed quantities. This is the
strategy followed by \cite{marian_dipolo12} to write the weak-field weak-line approximation for a 
non-rotating star in terms of the observed fluxes (and the parameters of the CLV and the Gaussian profile). So:
\begin{eqnarray}
F'_{I}(\lambda)&=& -\left(\psi \overline{f\mathcal{H}'} + \psi'
\overline{f\mathcal{H}}\right) \nonumber \\
F''_{I}(\lambda)&=& -\left(\psi \overline{f\mathcal{H}''} + \psi''
\overline{f\mathcal{H}} +2\psi'\overline{f\mathcal{H}'}\right),
\end{eqnarray}
which gives, after isolation of $\psi(\lambda)$, $\psi'(\lambda)$ and
$\psi''(\lambda)$:
\begin{eqnarray}
F_{Q}(\lambda) &=& -\frac{{C}^2}{4} \lambda_0^4 G \left[
\frac{\overline{(B_x^2-B_y^2) f \mathcal{H}}}{\overline{f\mathcal{H}}}
F''_{I}
+2\left( \frac{\overline{(B_x^2-B_y^2) f \mathcal{H}'}}{\overline{f\mathcal{H}}}
- \frac{\overline{(B_x^2-B_y^2) f
\mathcal{H}}~\overline{f\mathcal{H}'}}{\overline{f\mathcal{H}}^2} \right)
F'_{I}\right. \nonumber \\
 &+& \left. \left( \frac{\overline{(B_x^2-B_y^2) f
\mathcal{H}''}}{\overline{f\mathcal{H}}} -
\frac{\overline{(B_x^2-B_y^2)f\mathcal{H}}~\overline{f\mathcal{H}''}}{\overline{
f\mathcal{H}}^2} 
+2\frac{\overline{(B_x^2-B_y^2)f\mathcal{H}}~\overline{f\mathcal{H}'}^2}{
\overline{f\mathcal{H}}^3} \right. \right. \nonumber \\
&-& \left. \left. 2\frac{\overline{(B_x^2-B_y^2)\mathcal{H}'}~\overline{f\mathcal{H}'}}{\overline
{f\mathcal{H}}^2} \right) (F_c-F_{I})\right] \nonumber \\
F_{U}(\lambda) &=& -\frac{{C}^2}{2} \lambda_0^4 G \left[
\frac{\overline{B_xB_y f \mathcal{H}}}{\overline{f\mathcal{H}}} F''_\mathcal{I}
+2\left( \frac{\overline{B_xB_y f \mathcal{H}'}}{\overline{f\mathcal{H}}} -
\frac{\overline{B_xB_y f
\mathcal{H}}~\overline{f\mathcal{H}'}}{\overline{f\mathcal{H}}^2} \right)
F'_{I}\right. \nonumber \\
 &+& \left. \left( \frac{\overline{B_xB_y f
\mathcal{H}''}}{\overline{f\mathcal{H}}} -
\frac{\overline{B_xB_yf\mathcal{H}}~\overline{f\mathcal{H}''}}{\overline{
f\mathcal{H}}^2} 
+2\frac{\overline{B_xB_yf\mathcal{H}}~\overline{f\mathcal{H}'}^2}{\overline{
f\mathcal{H}}^3}
-2\frac{\overline{B_xB_y\mathcal{H}'}~\overline{f\mathcal{H}'}}{\overline{
f\mathcal{H}}^2} \right) (F_c-F_{I})\right] \nonumber \\
F_{V}(\lambda) &=& -{C} \lambda_0^2 g \left[ \frac{\overline{B_z
f \mathcal{H}}}{\overline{f\mathcal{H}}} F'_{I}  + 
\left( \frac{\overline{B_z f
\mathcal{H}}~\overline{f\mathcal{H}'}}{\overline{f\mathcal{H}}^2} - \frac{\overline{B_z f \mathcal{H}'}}{\overline{f\mathcal{H}}}\right)
(F_c-F_{I}) \right]
\label{eq:fI_star_observer}
\end{eqnarray}
These expressions are the generalization of the weak-field weak-line approximation to an unresolved
star with an arbitrary velocity and magnetic field on the surface. The dependency on the
local absorption $\eta$ has disappeared once the observed $F_{I}$ is used. However,
the local broadening is still present and has to be either inferred from the observations or fixed
to a reasonable value.
The set of Eqs. (\ref{eq:fI_star_observer}) represent a description of the Stokes parameters 
in terms of observed quantities and also in terms of wavelength-dependent unknowns 
that are explicit functions of the field and velocity distribution on the star
(e.g., $\overline{B_z f \mathcal{H}} / \overline{f H}$). Both contributions are 
conveniently separated. In principle, one could use these expressions to obtain
information about the magnetism of the star following two strategies. The first one is to parameterize
these terms by fixing a model for the magnetic structure and infer the values of
the parameters. This is the approach we follow in the next section. The second one is
to non-parametrically infer these wavelength-dependent unknowns directly using
a suitable inversion procedure. Since this would require additional regularization
techniques, a full treatment of the problem would have to be done in a Bayesian framework.

An interesting property of the integrated Stokes parameters when rotation is taken into account
is that the circular polarization flux depends on the derivative with wavelength of the flux ($F_{I}'$) 
but also on the flux itself ($F_{I}$). Likewise, the linear polarization fluxes depend on
the flux and its first and second order derivative with respect to the wavelength.
When particularizing the polarized flux of
Eq. (\ref{eq:fI_star_observer}) to the non-rotating star, we recover the equations developed
by \cite{marian_dipolo12}. This is a consistency check of our equations.

\section{The rotating magnetic dipole}
\label{sec:dipole}
\begin{figure*}[!t]
\includegraphics[width=\textwidth]{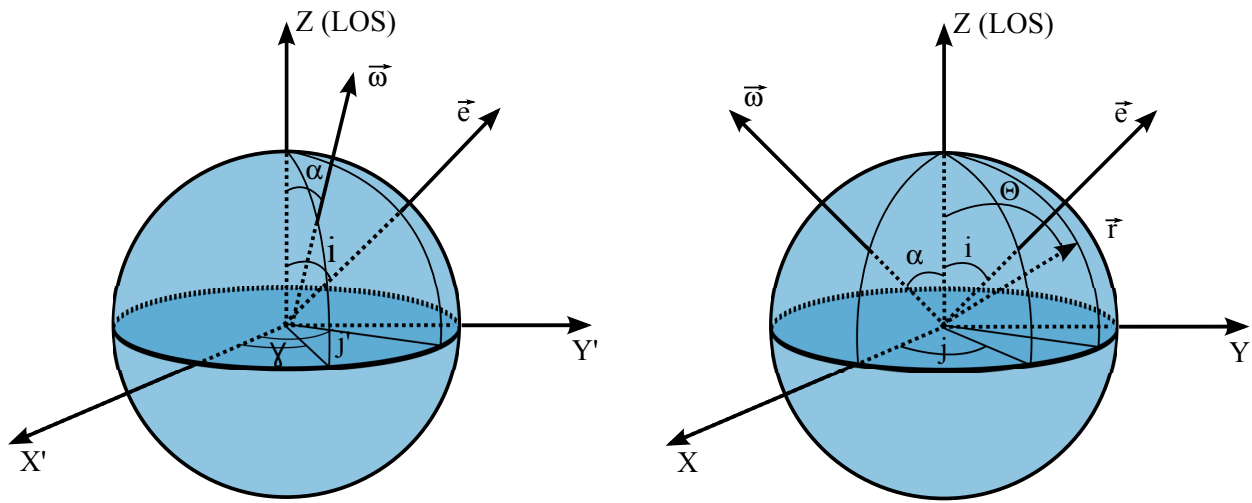}
\caption{This figure shows the two reference systems used in this work. We consider that the observer
is along the $Z$ axis, that is shared between both reference systems. The reference system $X'Y'Z$ is
an arbitrary one defined by the observer, where $X'$ defines the reference direction for $Q>0$. In order
to simplify the computations, we use the $XYZ$ reference system where the rotation angular velocity
vector is in the $XZ$ plane. The transformation between the two reference systems is given by a 
standard rotation with an angle $\gamma$. The vector $\mathbf{e}$ defines the dipolar magnetic field, while
$\mathbf{r}$ defines the position on the stellar surface.}
\label{fig:fig_geom_dipole}
\end{figure*}

Let us consider a star that is rotating with an angular velocity vector $\omegabold$, as shown in Fig. \ref{fig:fig_geom_dipole}.
In this section we use two different reference systems for our computations. The arbitrary reference system
$X'Y'Z$, shown in the left panel of Fig. \ref{fig:fig_geom_dipole}, is the one associated with an 
observer. The line-of-sight (LOS) is along the $Z$ axis and the axis $X'$ (arbitrarily chosen by the
observer) defines the reference direction for positive Stokes $Q$. The angular velocity vector is 
defined in this reference system by its modulus ($\omega$), 
the inclination angle ($\alpha$) and the azimuth ($\gamma$) with respect to the $X'$ axis.
The second reference system, $XYZ$, is obtained after a rotation of an angle $\gamma$ around the
$Z$ axis, so that the angular velocity vector is now contained in the $XZ$ plane.
This second reference system will help us in obtaining the final analytical expressions.

One of the simplest non-trivial configuration of a stellar magnetic field is that of a dipole, whose axis is determined
by a unit vector $\mathbf{e}$ along the dipole moment and an intrinsic dipolar field. The three components of the
unit vector $\mathbf{e}$ on the cartesian system $XYZ$ are:
\begin{eqnarray}
e_x&=&\sin{i}\cos{j} \nonumber \\
e_y&=&\sin{i}\sin{j} \nonumber \\
e_z&=&\cos{i},
\end{eqnarray}
where $i$ is the inclination of the dipolar moment with respect to the LOS and $j$ is its 
azimuth with respect to the $X$ axis. In the
arbitrary reference system $X'Y'Z$, the azimuth is transformed to $j'=j+\gamma$. The
emergent Stokes parameters $Q(\lambda)$ and $U(\lambda)$ are then transformed to the arbitrary
reference system following Eq. (\ref{eq:rotation_stokes}). Note
that our notation can be translated to that used by \cite{landolfi93} making the
transformations (in format ours$\to$theirs): $\alpha \to i$, $j' \to \Lambda$, $\gamma \to \Theta$, $i \to l$. Therefore,
the cosine of the inclination of the dipole axis with respect to the rotational axis, usually
termed $\beta$, is plainly given by the dot product $\omegabold \cdot \mathbf{e}$, i.e., 
$\cos \beta = \sin \alpha \sin i \cos j' + \cos \alpha \cos i$.

The magnetic 
field at the surface of the star of a dipole placed at the center of the star can be expressed 
in cartesian coordinates in the $XYZ$ system as:
\begin{eqnarray}
B_x&=&-\frac{B_p}{2}\left[ e_x - 3x(x e_x+y e_y+z e_z) \right] \nonumber \\
B_y&=&-\frac{B_p}{2}\left[ e_y - 3y(x e_x+y e_y+z e_z) \right] \nonumber \\
B_z&=&-\frac{B_p}{2}\left[ e_z - 3z(x e_x+y e_y+z e_z) \right],
\end{eqnarray}
where $B_p$ is the intensity of the magnetic field at the poles. An arbitrary
point $(x,y,z)$ in the visible surface of the star can be written in terms of the polar coordinates on
the plane of the sky $\Sigma$ as:
\begin{eqnarray}
x &=& \rho \cos \theta \nonumber \\
y &=& \rho \sin \theta \nonumber \\
z &=& \sqrt{1-\rho^2},
\end{eqnarray}
where we have selected the positive solution for $z$ to isolate the visible half of the star.

For simplicity, we assume that the star is rotating as a solid body because it greatly
simplifies the equations (the introduction of a non-rigid rotation is formally identical, except
for the computation of the $X_i(n)$ integrals), so that:
\begin{equation}
\mathbf{v}=\omegabold \times \mathbf{r}.
\end{equation}
In terms of the polar coordinates on the stellar surface, the component of the velocity along the LOS
in the $XYZ$ reference system is given by:
\begin{equation}
v_z=\omega \rho \sin{\alpha} \sin{\theta}. 
\end{equation}

Taking all these definitions into account and after some tedious algebra, it is possible
to find relatively simple expressions for the $X_i(n)$ integrals, which now depend
on the set of parameters $\{i,j,\alpha,\omega,n,a,b\}$. The procedure is to plug
the functional form of the dipolar magnetic field and carry out the
angular and radial integrals on the visible disk. We end up with:
\begin{eqnarray}
X_1(i,j,\alpha,\omega,n,a,b) &=& (\omega \sin \alpha)^n \left\{ A_1(i,j,\alpha,\omega,n) \left[ (1-a-b) F_1(n) + a F_2(n) + b
F_3(n)\right] \right.\nonumber \\
&+& B_1(i,j,\alpha,\omega,n) \left[ (1-a-b) F_3(n) + a F_4(n) + b F_5(n)\right] \nonumber \\
&+& C_1(i,j,\alpha,\omega,n) \left[ (1-a-b) F_2(n+1) + a F_3(n+1) + b
\left. F_4(n+1)\right] \right\} \nonumber \\
X_2(i,j,\alpha,\omega,n,a,b) &=& (\omega \sin \alpha)^n \left\{A_2(i,j,\alpha,\omega,n) \left[ (1-a-b) F_1(n+2) + a F_2(n+2) + b
F_3(n+2)\right] \right. \nonumber \\
&+& B_2(i,j,\alpha,\omega,n) \left[ (1-a-b) F_1(n+4) + a F_2(n+4) + b
F_3(n+4)\right] \nonumber \\
&+& C_2(i,j,\alpha,\omega,n) \left[ (1-a-b) F_2(n+1) + a F_3(n+1) + b
F_4(n+1)\right] \nonumber \\
&+& D_2(i,j,\alpha,\omega,n) \left[ (1-a-b) F_2(n+3) + a F_3(n+3) + b
F_4(n+3)\right] \nonumber \\
&+& \left. E_2(i,j,\alpha,\omega,n) \left[ (1-a-b) F_1(n) + a F_2(n) + b F_3(n)\right] \right\} \nonumber \\
X_3(i,j,\alpha,\omega,n,a,b) &=& (\omega \sin \alpha)^n \left\{ A_3(i,j,\alpha,\omega,n) \left[ (1-a-b) F_1(n+2) + a F_2(n+2) + b
F_3(n+2)\right] \right. \nonumber \\
&+& B_3(i,j,\alpha,\omega,n) \left[ (1-a-b) F_1(n+4) + a F_2(n+4) + b
F_3(n+4)\right] \nonumber \\
&+& C_3(i,j,\alpha,\omega,n) \left[ (1-a-b) F_2(n+1) + a F_3(n+1) + b
F_4(n+1)\right] \nonumber \\
&+& D_3(i,j,\alpha,\omega,n) \left[ (1-a-b) F_2(n+3) + a F_3(n+3) + b
F_4(n+3)\right] \nonumber \\
&+& \left. E_3(i,j,\alpha,\omega,n) \left[ (1-a-b) F_1(n) + a F_2(n) + b F_3(n)\right] \right\} \nonumber \\
X_4(i,j,\alpha,\omega,n,a,b) &=& (\omega \sin \alpha)^n A_4(i,j,\alpha,\omega,n) \left[ (1-a-b) F_1(n) + a F_2(n) + b F_3(n)\right].
\label{eq:X_functions}
\end{eqnarray}
These functions depend on the orientation of the dipole given by
the angles $i$ and $j$, the angular velocity parameters and the
parameters of the CLV and the index $n$ of the summation. The
following functions take into account the dependence on the
orientation of the dipole and only depend on angular integrals:
\begin{eqnarray}
A_1(i,j,\alpha,\omega,n) &=&  S(n) \cos i \nonumber \\
B_1(i,j,\alpha,\omega,n) &=& -3  S(n) \cos i \nonumber \\
C_1(i,j,\alpha,\omega,n) &=& -3 S(n+1) \sin i \sin j \nonumber \\
A_2(i,j,\alpha,\omega,n) &=& 3  \cos^2 i \left[3 S(n)-8S(n+2) \right] -6\sin^2 i \cos^2 j
S(n) + 6 S(n+2) \nonumber \\
B_2(i,j,\alpha,\omega,n) &=& 9 \cos^2 i \left[ S(n+2) - S(n) + 2S(n+4) \right] \nonumber \\
&+& 9 \sin^2
i \cos^2 j \left[ S(n) - 4 S(n+2) + 4S(n+4)\right] \nonumber \\
&+& 9S(n+2) - 18S(n+4)  \nonumber \\
C_2(i,j,\alpha,\omega,n) &=& 6 \cos i \sin i \sin j S(n+1) \nonumber \\
D_2(i,j,\alpha,\omega,n) &=& 18  \cos i \sin i \sin j \left[ S(n+1) - 2S(n+3)
\right]  \nonumber \\
E_2(i,j,\alpha,\omega,n) &=& \sin^2 i \cos (2j) S(n) \nonumber \\
A_3(i,j,\alpha,\omega,n) &=& -3  \sin^2 i \cos j \sin j S(n) \nonumber \\
B_3(i,j,\alpha,\omega,n) &=& 18  \sin^2 i \cos j \sin j \left[ S(n+2) - S(n+4) \right]  \nonumber \\
C_3(i,j,\alpha,\omega,n) &=& -3 \cos i \sin i \cos j S(n+1) \nonumber \\
D_3(i,j,\alpha,\omega,n) &=& 18 \cos i \sin i \cos j \left[ S(n+1) - S(n+3) \right] \nonumber \\
E_3(i,j,\alpha,\omega,n) &=& \sin^2 i \cos j \sin j S(n) \nonumber \\
A_4(i,j,\alpha,\omega,n) &=&  S(n),
\label{eq:ABCDE_functions}
\end{eqnarray}
with the angular integral $S(n)$ given by:
\begin{equation}
S(n) = \int_0^{2\pi} d\theta \sin^n \theta = \left[ 1+(-1)^n
\right] \sqrt{\pi} \Gamma \left( \frac{1+n}{2} \right) \Gamma^{-1}
\left( 1+\frac{n}{2} \right),
\label{eq:S_functions}
\end{equation}
which is valid for $n>-1$. Likewise, the radial integrals $F_i(n)$
are independent on the
orientation of the dipole and are computed from:
\begin{eqnarray}
F_1(n) &=& \int_0^1 \rho^{n+1} d\rho = \frac{1}{2+n} \nonumber \\
F_2(n) &=& \int_0^1 \sqrt{1-\rho^2} \rho^{n+1} d\rho =
\frac{\sqrt{\pi}}{4} \Gamma \left( 1+\frac{n}{2} \right) \Gamma^{-1}
\left( \frac{5+n}{2} \right) \nonumber \\
F_3(n) &=& \int_0^1 (1-\rho^2) \rho^{n+1} d\rho = \frac{1}{8+6n+n^2}
\nonumber \\
F_4(n) &=& \int_0^1 (1-\rho^2) \sqrt{1-\rho^2} \rho^{n+1} d\rho =
\frac{3\sqrt{\pi}}{8} \Gamma \left( 1+\frac{n}{2} \right) \Gamma^{-1}
\left( \frac{7+n}{2} \right) \nonumber \\
F_5(n) &=& \int_0^1 (1-\rho^2)^2 \rho^{n+1} d\rho = \frac{8}{(2+n)(4+n)(6+n)},
\label{eq:F_functions}
\end{eqnarray}
which are valid if $n>-2$.

Once the expressions for $X_i(n)$ have been particularized to the case of a rotating dipole,
it is possible to analyze in detail the shape of the profiles according to Eq. (\ref{eq:fI_star}). 
We analyze the symmetry properties of the circular polarization profile, although this can be extended easily to the linear
polarization profiles. Recall that the profile $\psi(\lambda)$ is a symmetric function of wavelength 
with respect to the rest wavelength of the line, while $\psi'(\lambda)$ is antisymmetric. 
According to the properties of the Hermite polynomials, the terms
with $n=0,2,4,\ldots$ in the expression for $\overline{B_z f \mathcal{H}}$ in Eq. (\ref{eq:averages}) are symmetric, while
those with $n=1,3,5,\ldots$ are always antisymmetric. Likewise, for $\overline{B_z f \mathcal{H}'}$, the terms with 
$n=0,2,4,\ldots$ are antisymmetric, while those with $n=1,3,5,\ldots$ are symmetric. Consequently, the polarized
flux $F_{V}(\lambda)$ can be written as the addition of two functions, one symmetric and the other
one antisymmetric:
\begin{equation}
F_{V}(\lambda) = F_{V}^\mathrm{sym}(\lambda) + F_{V}^\mathrm{antisym}(\lambda),
\end{equation}
where
\begin{eqnarray}
F_{V}^\mathrm{sym}(\lambda) &=& 
{C}\lambda_0^2g \left[ \psi'(\lambda) \overline{B_zf\mathcal{H}}_\mathrm{antisym} +\psi(\lambda) \overline{B_zf\mathcal{H}'}_\mathrm{sym} \right] \nonumber \\
F_{V}^\mathrm{antisym}(\lambda) &=& 
{C}\lambda_0^2g \left[ \psi'(\lambda) \overline{B_zf\mathcal{H}}_\mathrm{sym} +\psi(\lambda) \overline{B_zf\mathcal{H}'}_\mathrm{antisym} \right].
\end{eqnarray}
Taking into account the definition of the $X_1(n)$ function of Eq. (\ref{eq:X_functions}), it is easy
to verify that the symmetric contribution to $F_{V}(\lambda)$ depends only on $\sin i \sin j$, while
the antisymmetric contribution depends only on $\cos i$. Consequently, we will find standard
antisymmetric circular polarization profiles whenever $i=0$, while they will be fully
symmetric when $i=90^\circ$.

Stellar rotation is taken into account by keeping fixed the LOS
and rotating the vector $\mathbf{e}$ around
the angular rotation velocity $\omegabold$. This is accomplished by computing the new rotated
dipolar moment unit vector as:
\begin{equation}
\mathbf{e}_\mathrm{rot} = \mathbf{R} \mathbf{e},
\end{equation}
where the rotation matrix is given by:
\begin{eqnarray}
\mathbf{R} =
\left[
\begin{array}{ccc}
\cos \phi+\sin^2 \alpha (1-\cos \phi) & -\cos \alpha \sin \phi & \sin \alpha \cos \alpha(1-\cos \phi)\\
\cos \alpha \sin \phi & \cos \phi  & -\sin \alpha \sin \phi \\
\sin \alpha \cos \alpha (1-\cos \phi) & \sin \alpha \sin \phi & \cos \phi + \cos^2 \alpha(1-\cos \phi)
\end{array}
\right],
\end{eqnarray}
with $\phi$ indicating the stellar rotation phase and taking values between $0$ and $2\pi$. Note
that our initial of phases is an arbitrary position of the vectors, while using
the standard notation, the initial phase coincides with the $\omegabold$ and $\mathbf{e}$
vectors lying in the same plane parallel to the $Z$ axis. This implies a phase shift
between our $\phi$ angle and the standard definition that is given by:
\begin{equation}
\tan \phi_0 = -\frac{\sin i \sin j}{\cos \alpha \sin i \cos j - \sin \alpha \cos i}.
\end{equation}
Finally, 
the inclination and azimuth angles with respect to the $XYZ$ reference system are obtained
using the standard relations, taking into account that $\mathbf{e}_\mathrm{rot}$ is a unit
vector:
\begin{equation}
i_\mathrm{rot} = \arccos \left[ (e_\mathrm{rot})_z \right], \qquad j_\mathrm{rot} = \arctan \frac{(e_\mathrm{rot})_y}{(e_\mathrm{rot})_x}
\end{equation}

\section{Technicalities}
All the functions needed in the previous equations are well behaved and can be
computed numerically without much complexity. The main difficulty resides on the
fact that, if the rotation velocity $\omega \sin \alpha$ is large, the
convergence
of the expansion in Eqs. (\ref{eq:averages_surface}) is slow and
oscillating, thus
making it hard to compute. The main problem is that the Hermite polynomials grow in
amplitude while the coefficients decrease slow. Only after many terms (the number of which depend on
the ratio between the rotation velocity and the broadening), the coefficients start to compensate
for the increase in the amplitude of the Hermite polynomials and convergence is eventually obtained. This
is a well-known defect of the Gram-Charlier expansion \citep[e.g.,][]{blinnikov98}.
When the rotation velocity is much larger than the Doppler width of the local spectral
line (larger than a factor $\sim 8$), the compensation between the amplitude increase of the Hermite polynomials and the 
coefficients of the series expansion occurs for very large values of $n$ [see Eqs. (\ref{eq:averages_surface})].
At this moment, the double precision computations start to become unprecise and the terms in the
expansion do not compensate adequately. This purely numerical drawback can be overcome if arbitrary
precision methods are used or if more elaborate methods like the Edgeworth expansion are adapted to our problem \citep{blinnikov98}.
In any case, we have been able to numerically compute expansions up to $\omega \sin\alpha \sim 8
\sigma c /\lambda_0$. In other words, our method with a double precision summation of the expansions
shown in Eqs. (\ref{eq:averages_surface}) is presently applicable for rotation velocities
not larger than 8 times the Doppler width of the local spectral line.

In any case, it is advantageous to use the following set of modified
Hermite polynomials:
\begin{equation}
\textit{He}_n^\star(x) = \left( \frac{\lambda_0}{\sigma c} \right)^n \frac{1}{n!} \textit{He}_n(x),
\end{equation}
which fulfill the following recursion formula:
\begin{equation}
\textit{He}_n^\star(x) = \left( \frac{\lambda_0}{\sigma c} \right) \frac{1}{n} x \textit{He}_{n-1}(x) - \left( \frac{\lambda_0}{\sigma c} \right)^2
\frac{1}{n} \textit{He}_{n-2}(x).
\end{equation}
Computing the Hermite polynomials with this recursion formula, the
expansions of Eqs. (\ref{eq:averages_surface})
can be extended to much larger values of $n$. The advantage of using the modified
$\textit{He}_n^\star(x)$ Hermite polynomials is that their amplitude grow much slower with $n$ than $\textit{He}_n(x)$ due to the 
inclusion of the term $\lambda_0/(\sigma c)$ and $n!$. This facilitates the 
computation of each term in the expansions without numerical overflows. We note that
the Hermite polynomials are computed using their recursion relations \citep{abramowitz72}.

From the computational point of view, once the
rotation angular velocity of the star is set, the wavelength dependence of
Eqs. (\ref{eq:averages_surface}) and their derivatives in terms of the Hermite
polynomials can be precomputed once and used in all subsequent calculations. This
greatly accelerates the process because only the $X_i(i,j,n)$ functions have to be
computed for a given orientation of the dipolar moment and added until convergence. 

\begin{figure*}
\includegraphics[width=0.5\textwidth]{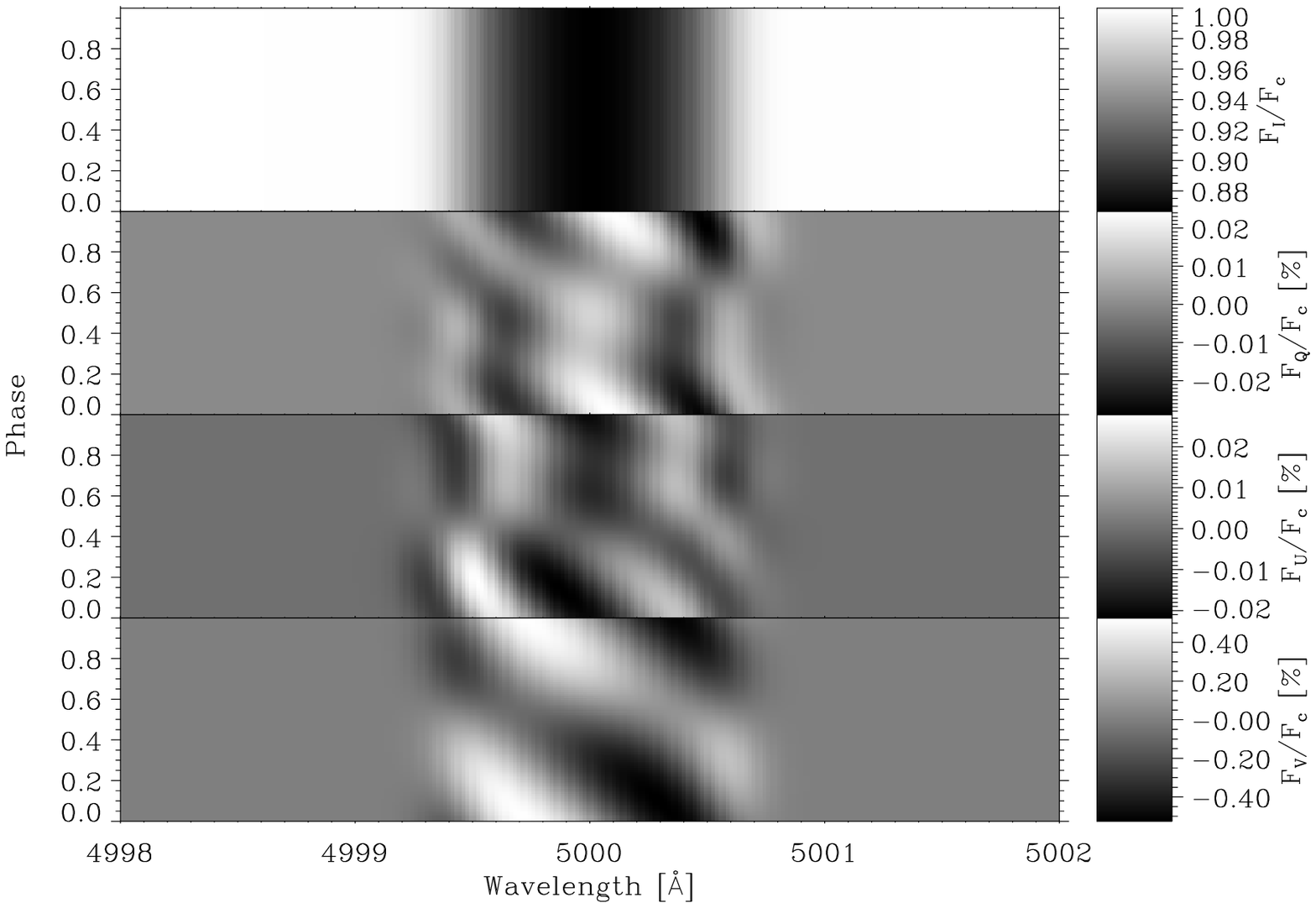}
\includegraphics[width=0.5\textwidth]{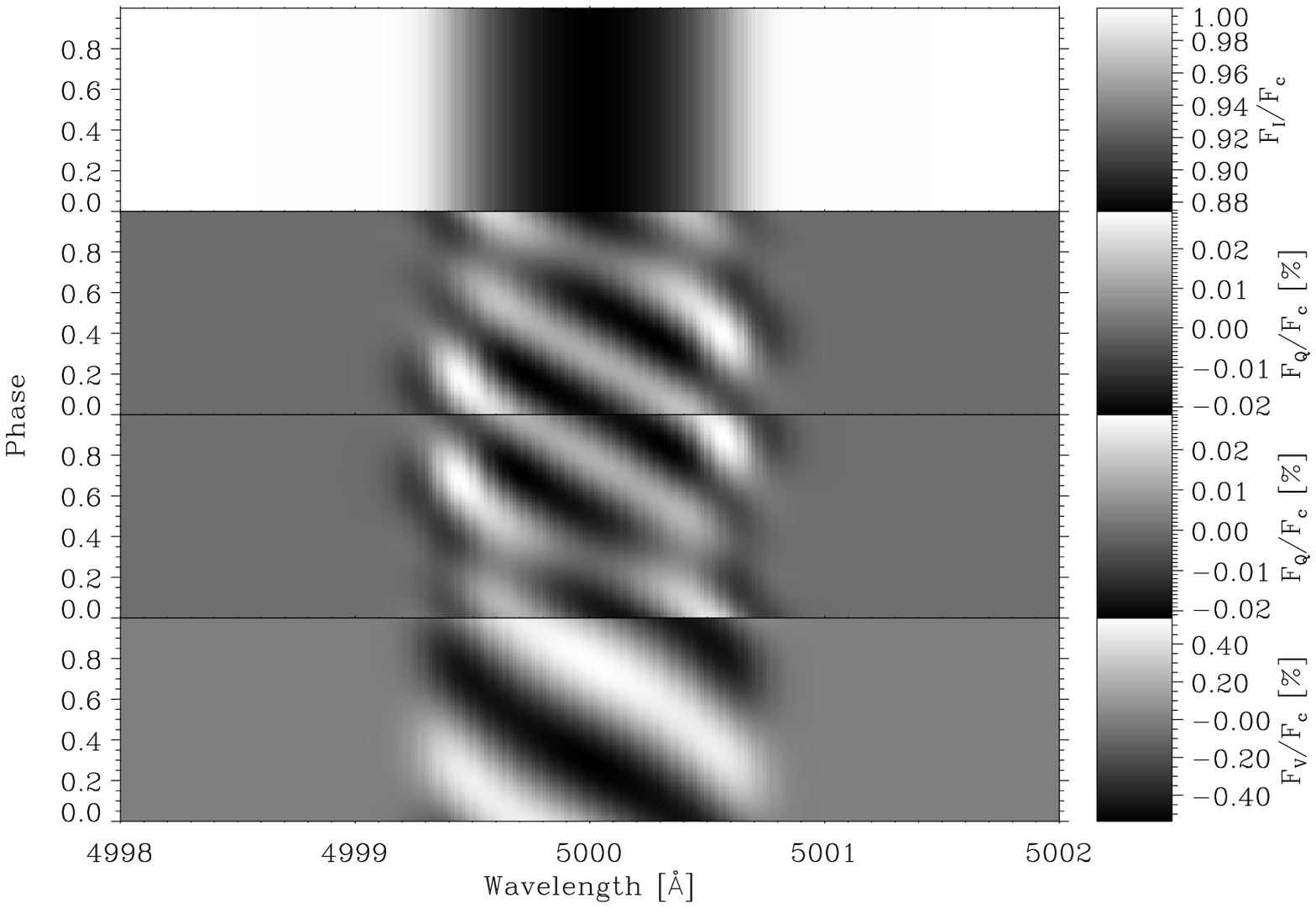}
\caption{Examples of the flux normalized to the continuum intensity flux for a full rotational phase. Both
examples share $a=0.2$, $b=0.2$, $B_p=500$ G and $\omega R_\mathrm{star}=40$ km s$^{-1}$, where
$R_\mathrm{star}$ is the stellar radius. The left panel corresponds
to the case $i=45^\circ$, $j'=80^\circ$, $\alpha=70^\circ$ and $\gamma=70^\circ$. The right panel
has $i=45^\circ$, $j'=30^\circ$, $\alpha=90^\circ$ and $\gamma=20^\circ$. All polarized fluxes are given in percentages.}
\label{fig:examples}
\end{figure*}

A free computer program in IDL is available to compute the observed Stokes fluxes for a rotating dipole
using the expressions in this paper\footnote{\texttt{http://www.iac.es/project/magnetism/dipole}}. In
this non-optimized version of the code, we synthesize profiles evaluated at 500 wavelength points
with a step of 8 m\AA\ in $\sim$35-40 ms.

\section{Illustrative examples}
Given our assumptions (weak-field, weak-line, fixed center-to-limb variation, global dipolar
field), one of the most direct applications of our equations is to study the magnetism
of hot stars. For this reason, we use typical values of the broadening and rotation velocity
of such stars \citep[e.g.,][]{petit_wade12}. We show in Fig.
\ref{fig:examples} two examples for a dipolar magnetic field with different field configurations.
Both computations share the parameters for the CLV ($a=0.2$ and $b=0.2$), the magnetic field
strength in the pole ($B_p=500$ G) and the rotation velocity on the surface ($v=40$ km s$^{-1}$).
The artificial selected spectral line is centered at $\lambda_0=5000$ \AA\ and has an effective Land\'e
factor of $g=3$ and its equivalent for linear polarization of $G=9$. Since the magnetic field 
takes its maximum value at the poles, the spectral line can be safely considered in the weak-field regime.
The line broadening is set
to 0.1 \AA, typical of an observation at a spectral resolution of $R=65000$ of a line with
a microturbulent width of $\sim$5 km s$^{-1}$. These values of the broadening and
rotation velocity are close to the numerical limit that we presently have for computing the 
summations of Eqs. (\ref{eq:averages}). Given this limitation,
our scheme is presently not applicable to fast rotators with narrow local absorption profiles. The computed intensity flux normalized
to the continuum flux is shown
in the upper panel of each figure. It is independent of the phase since, under the weak field
regime, the intensity does not depend on the magnetic field and we have assumed that the star
is structureless. However, it takes into account the rotation broadening and the presence
of a CLV. Concerning the linear polarization fluxes, we see maximum amplitudes around 0.02\%, while
this increases to values around 100 times larger for circular polarization. This is a direct consequence of
the fact that linear polarization appears only in second order.

\section{Conclusions}
We have generalized the expressions for the Stokes parameters in the weak-field weak-line regime to the
stellar case in which rotation is taken into account. This produces an intensity flux profile
that is broadened by rotation while it is insensitive to the magnetic field. As a consequence,
our formalism can only be applied to stars in which the intensity profile is not observed to
be modulated by rotation. However, the circular and linear polarized fluxes are sensitive to
the topology of the magnetic field. We have demonstrated that the inclusion of Doppler shifts
generates a dependence of circular polarization of the intensity flux itself, and not only
on its wavelength derivative. Likewise, the linear polarization profiles depend now on the
first and second wavelength derivative and on the intensity flux itself. We have verified also
that the expressions developed in this paper simplify to those obtained by \cite{marian_dipolo12}
in the non-rotating dipole case. The general weak-field expressions have been particularized to a rotating dipole. The extension 
to a more general multipolar magnetic field is relatively easy and can be done fast if analytical
expressions for the angular and radial integrals exist. In subsequent papers of this series
we will deal with thin magnetized disks in Keplerian rotation and unresolved structures
in the quiet Sun. Likewise, we will also explore the possibility of using
more elaborate expansions to improve the convergence.

\begin{acknowledgements}
We thank R. Manso Sainz for fruitful discussions.
Financial support by the 
Spanish Ministry of Economy and Competitiveness through projects AYA2010-18029 (Solar Magnetism and Astrophysical 
Spectropolarimetry) and Consolider-Ingenio 2010 CSD2009-00038 is gratefully acknowledged.
\end{acknowledgements}


\end{document}